\documentclass{article}
\usepackage[T1]{fontenc} 
\usepackage[utf8]{inputenc} 
\usepackage{ismir,amsmath,cite,url}
\usepackage{graphicx}

\usepackage{color}

\usepackage[firstpage]{draftwatermark}
\definecolor{lightgray}{rgb}{0.9,0.9,0.9}
\definecolor{darkgray}{rgb}{0.4,0.4,0.4}
\SetWatermarkFontSize{12pt}
\SetWatermarkScale{1.1}
\SetWatermarkAngle{90}
\SetWatermarkHorCenter{202mm}
\SetWatermarkVerCenter{170mm}
\SetWatermarkColor{darkgray}
\SetWatermarkText{Late-Breaking / Demo Session Extended Abstract, ISMIR 2024 Conference}

\def\lbd{}	

\title{Understanding Human Perception of Music Plagiarism Through a Computational Approach}

\twoauthors
  {Daeun Hwang} {University of California, Santa Cruz\\  {\tt dhwang8@ucsc.edu}}
  {Hyeonbin Hwang} {Korea Advanced Institute of Science and Technology\\ {\tt hbin0701@kaist.ac.kr}}

\def\authorname{D. Hwang and H. Hwang}

\begin{document}

\maketitle
\begin{abstract}
There is a wide variety of music similarity detection algorithms, while discussions about music plagiarism in the real world are often based on audience perceptions. Therefore, we aim to conduct a study to examine the key criteria of human perception of music plagiarism, focusing on the three commonly used musical features in similarity analysis: melody, rhythm, and chord progression. After identifying the key features and levels of variation humans use in perceiving musical similarity, we propose a LLM-as-a-judge framework that applies a systematic, step-by-step approach, drawing on modules that extract such high-level attributes.
\end{abstract}
\section{Introduction} 
Plagiarism in music has been a constant issue over time. Not only could it hurt the reputation of musicians, but it may challenge and question the originality and creativity in music \cite{li2024textual, deprisco2017plagiarism}. While there are many attempts to tackle music similarity, there have been discrepancies in music similarity detection algorithms since there is a lack of certain standards \cite{cason2012singing, flexer2014inter}. Furthermore, most existing works on music similarity have slightly different directions: many studies use content-based music analysis \cite{mcfee2012learning, foote1997content, knees2013survey}, or rather focus direction as for music recommendation \cite{shao2009music, han2018music}. While there are many different algorithms for music similarity detection, only human experts can possibly evaluate such complex multimodal matter comprehensively \cite{mullensiefen2009court}. Most of the existing discussion around plagiarism in the media or community revolves around and from the audience's perception. Therefore, this work aims to understand how people perceive music similarity from a computational viewpoint. We will further define data-driven criteria of plagiarism based on different weights of musical elements affecting human perception towards music similarity to propose a model to predict potential plagiarism cases.

\section{Related work}
Existing literature demonstrates a wide range of variety in defining and measuring music similarity. Many studies use multimodal approaches, often involving multiple level or space analysis \cite{shiu2005musical, bogdanov2011unifying} or visualization-based analysis \cite{simonetta2018symbolic, deprisco2017plagiarism}. Existing studies commonly used melody \cite{li2024textual, mullensiefen2004cognitive, casey2006importance}, rhythm \cite{ellis2008cross, holzapfel2008rhythmic, pohle2009rhythm, cao2012similarity}, and chord progression \cite{dehaas2008tonal, dehaas2011comparing, wongsaroj2014music}. Melody is the key concept in musicology that determines the main theme of the song. It is often analyzed as pitch or in combination with rhythm. In our study, a melodic pattern will be viewed as a sequence of notes with different pitches shown through relative pitch representation in semitones \cite{downie2003music} instead of absolute pitch. Rhythmic characteristics in context of music similarity can be defined as the temporal sequence of multiple notes and meter \cite{cao2012similarity}. According to Peeters, rhythmic features are mainly analyzed based on one of the five approaches: similarity matrix, features, temporal pattern, normalized periodicity, and source separation \cite{peeters2010spectral}. Commonly, it involves Dynamic Periodicity Warping (DPW) measures to distinguish rhythmic similarity since it reduces oversight of missing rhythmically similar pieces that have different tempo \cite{holzapfel2008rhythmic}. Since each chord possesses an impression of consonance or dissonance \cite{aiba2012chord}, tonality characters and combination of chords can form its own mood. Thus, chord progression evaluation for music similarity will be the summed chord similarity calculated from comparison of each pair of chords in two sequences \cite{wongsaroj2014music}. Although timbre is widely used to analyze music similarity \cite{bogdanov2011unifying, shiu2005musical, deprisco2017plagiarism}, it will not be included in this study, as the focus is on structural elements of the music rather than the overall sound characteristic or quality. Since timbre refers to the character of sound that distinguishes different instruments \cite{loughran2008mel, fales2002paradox}, this study will ensure that all test tracks use the same instrumental sound for consistency.

\section{Study on Human Perception of Similarity}
We will first conduct a study to gather data on how human listeners perceive music similarity. It will take place in the form of an online survey, where participants will listen to embedded audio files in each question and respond to the prompts. Participants will be recruited through convenient sampling, including individuals with both musical and non-musical experience. Any participants aged 18 or older with no difficulties in hearing abilities will be eligible to participate in the study. An informed consent will be provided to all participants prior to participation. 
We will create synthetic musical dataset with varying degrees of the commonly used criteria from previous studies on music similarity and plagiarism detection: melody, rhythm, and chord progression. Instead of using existing songs, generated music tracks will be used to maintain control over the tested variables, thus preventing confounding variables. Also, it will reduce copyright issues and biases from listeners’ prior knowledge or preferences. The generated tracks will be created using Logic Pro, a MIDI software. With the help of experts in the musical fields, we will first generate 5 tracks of approximately 30 seconds each with different chord progressions, melodies, and rhythm. All tracks will be recorded using the same default sound of the grand piano. Then, these original tracks will be processed with music21 \cite{cuthbert2010music21}, a python library for musical scripting. To create altered version, we will use variation levels of approximately 0\%, 30\%, 50\%, and 100\% to create distinct combinations of feature variations. The all-original and fully altered combinations will be excluded, resulting in a total of 62 variations. For melody variation, randomly selected notes will be altered from the original track using the transpose function. For chord progressions, several techniques will be applied, such as chord inversions, chord extensions, and substitute chords. For rhythmic variations, notes durations and syncopation will be altered. The tempo of the modified tracks will be adjusted as well.

Participants will be presented to 15 randomly selected pairs of music tracks. They will be asked to indicate whether they perceive the tracks as ‘plagiarized,’ rate the similarity between the tracks as a percentage, and provide a short comment on their reasoning. Participants’ responses will be analyzed to determine which feature and degree of variation had the most significant effects. In addition, the tracks will be analyzed using feature extractions such as mel-frequency cepstral coefficients (MFCC) and chroma feature to identify if there are any significant correlation with the participants’ perceptions.

\section{Prediction Model}
After defining any statistically significant criteria on musical features and the degree of similarity that influences listeners’ perceived music similarity in the context of plagiarism, we will train a model to successfully predict these results. As noted by previous research, many current music similarity systems rely on low-level features, which can result in biased outcomes \cite{poltronieri2023knowledge}. In light of this, we propose a novel framework based on the LLM-as-a-judge approach \cite{zheng2023judging} with specialized musical knowledge. Acting as a central intelligence, LLM would leverage high-level extracted features from other modules like chord progression, melody, and rhythm. Then, it will approach the evaluation of musical similarity with step-by-step critical thinking, similar to how human listeners systematically assess music using key elements and standards.

\section{Future Work}
In order to overcome the inconsistency and diversity in music similarity detection approaches, this work will verify significant patterns in listener perceptions towards music plagiarism. In future work, we will also thoroughly review and potentially categorize existing music similarity detection algorithms. By analyzing listeners’ criteria recorded through surveys, we will examine specific criteria and extent of variation of musical features that listeners consider when determining plagiarism. Using an LLM-as-a-judge approach, a prediction model for similarity evaluation will be developed based on the criteria analyzed in this study. In the following works, an LLM-as-a-judge model utilizing both low-level features and high-level features may be considered for a more accurate prediction. Through this work, we aim to expand understanding of human perception of music similarity and provide a model that facilitates detecting plagiarism risks.

\bibliography{ISMIRtemplate}

%
%
%
%
%

\end{document}